\documentclass[12pt,a4paper,final]{article}

\usepackage{amsmath}
\usepackage{amssymb}
\usepackage{amsthm}
\usepackage{exscale}
\usepackage[mathscr]{eucal}
\usepackage[normalem]{ulem}
\usepackage{floatrow}
\usepackage{bm}
\usepackage{graphicx}
\usepackage{color}
\usepackage{dsfont} 
\usepackage{lineno}
\usepackage{capt-of}
\usepackage{caption}
\usepackage{parskip}
\usepackage{diagbox} 

\usepackage[utf8]{inputenc} 

\usepackage{geometry}
\usepackage[dvipsnames]{xcolor}
\usepackage{tabularx}
\usepackage{amsmath}
\usepackage[sort&compress, numbers, square]{natbib}

\usepackage[%
    colorlinks=true,
    citecolor = gray,
    linkcolor= blue,
    urlcolor=gray
]{hyperref}

\usepackage{blindtext}
\usepackage{tcolorbox}
\definecolor{blizzardblue}{rgb}{0.67, 0.9, 0.93}
\definecolor{bisque}{rgb}{1.0, 0.89, 0.77}
\definecolor{bittersweet}{rgb}{1.0, 0.44, 0.37}
\definecolor{champagne}{rgb}{0.97, 0.91, 0.81}
\definecolor{deeppeach}{rgb}{1.0, 0.8, 0.64}
\definecolor{desertsand}{rgb}{0.93, 0.79, 0.69}
\definecolor{goldenpoppy}{rgb}{0.99, 0.76, 0.0}
\definecolor{aliceblue}{rgb}{0.94, 0.97, 1.0}
\definecolor{babyblueeyes}{rgb}{0.63, 0.79, 0.95}

\makeatletter

\title{\bf{Neural timescales from a computational perspective}}

\author{Roxana Zeraati$^{1,2,3,4,*}$, Anna Levina$^{1,2,3,4}$, Jakob H. Macke$^{3,4,5,6}$, Richard Gao$^{3,4,5,7,*}$
    \\ {\small $^1$ Max Planck Institute for Biological Cybernetics, T\"ubingen, Germany}
    \\ {\small $^2$ Self-Organization and Optimality in Neuronal Networks, University of T\"ubingen, T\"ubingen, Germany}
    \\ {\small $^3$ T\"ubingen AI Center, T\"ubingen, Germany }
    \\ {\small $^4$ Bernstein Center for Computational Neuroscience T\"ubingen, T\"ubingen, Germany}
    \\ {\small $^5$ Machine Learning in Science, University of T\"ubingen, T\"ubingen, Germany} 
    \\ {\small $^6$ Max Planck Institute for Intelligent Systems, T\"ubingen, Germany}
    \\{\small $^7$ Institute of Computer Science, Goethe University Frankfurt, Frankfurt, Germany}
    \\ {\small $^*$ These authors contributed equally to this work.}
    \\ {\small Correspondence: r.dg.gao@gmail.com, research@roxanazeraati.org }
}

\date{}

\begin{document}
\maketitle

\renewcommand{\figurename}{\bf{Fig.}}
\newcommand{\fig}{Fig.~}
\newcommand{\supsec}{Supplementary Sec.}
\newcommand{\supfig}{Supplementary Figure~}
\newcommand{\equ}{Eq.~}
\newcommand{\draft}{\color{red}}

\section*{Abstract} 
Neural activity fluctuates over a wide range of timescales within and across brain areas. 
Experimental observations suggest that diverse neural timescales reflect information in dynamic environments. 
However, how timescales are defined and measured from brain recordings vary across the literature. 
Moreover, these observations do not specify the mechanisms underlying timescale variations, nor whether specific timescales are necessary for neural computation and brain function. 
Here, we synthesize three directions where computational approaches can distill the broad set of empirical observations into quantitative and testable theories:
We review (i) how different data analysis methods quantify timescales across distinct behavioral states and recording modalities, (ii) how biophysical models provide mechanistic explanations for the emergence of diverse timescales, and (iii) how task-performing networks and machine learning models uncover the functional relevance of neural timescales.
This integrative computational perspective complements experimental investigations, providing a holistic view on how neural timescales reflect the relationship between brain structure, dynamics, and behavior.

\section{Introduction} 

Neural activity unfolds across a wide range of timescales.
These timescales are characterized by the decay rate of a neural signal's autocorrelation function. While precise definitions differ, recent experimental studies of neural timescales present a consistent picture across recording modalities, species, and cognitive tasks: timescales increase along the cortical hierarchy, are diverse within each brain area, and are often correlated with behavioral variables during cognitive tasks~\cite{cavanagh_diversity_2020,soltani_timescales_2021,wolff_intrinsic_2022}. 
As such, this quantity---the characteristic time constant of neuronal dynamics---may be an important signature of neural circuits across the brain. In particular, neural timescales have been proposed to play a ``representational'' role:  their diversity across the brain mirrors the multiple fluctuation timescales of behaviorally relevant information in a dynamic environment, enabling perception, memory, planning, and action~\cite{cavanagh_diversity_2020,soltani_timescales_2021,wolff_intrinsic_2022}. However, while a large body of experimental observations demonstrates the relevance of neural timescales, important differences exist between how these timescales are defined and measured.
Furthermore, it is experimentally challenging to probe the mechanisms underlying how neural timescales arise, and their precise role in the representation and transformation of information in the brain.

\begin{figure}[!t]
    \centering
     \includegraphics[trim=0 0 0 0, clip, width = \linewidth]{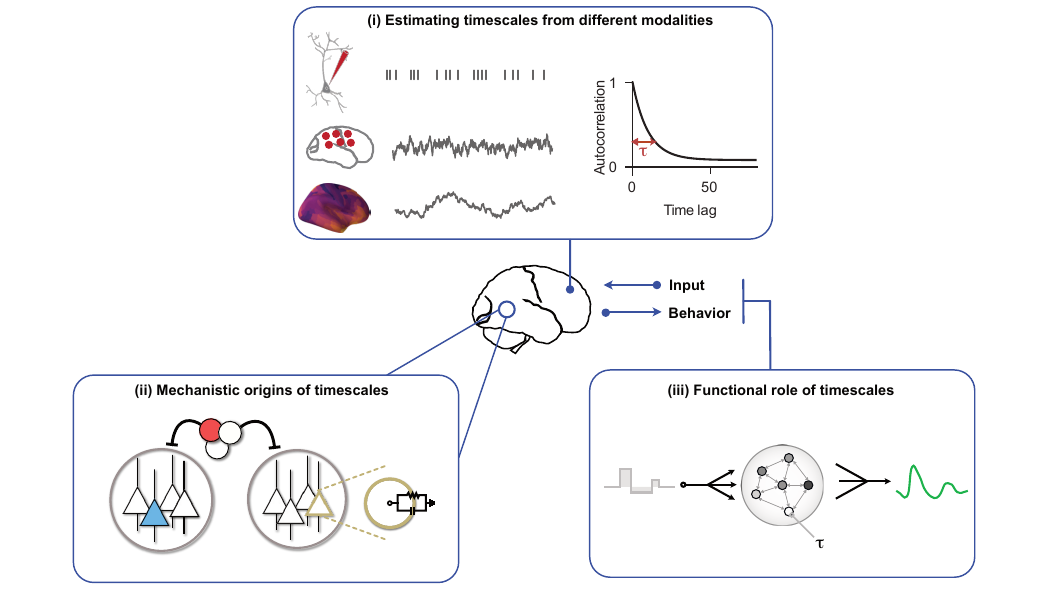} 
\caption{\textbf{Graphical summary.} This review discusses three main directions, demonstrating conceptually (i) how estimation methods contribute to advancing the measurement of timescales from different experimental paradigms (e.g., recording modalities), (ii) how mechanistic models can help in uncovering the underlying circuit origins of different timescales, and (iii) how functional modeling facilitates investigation of their computational relevance.
}
    \label{fig:intro_schem}
\end{figure}

Supplementing experimental investigations, we provide a complementary perspective by reviewing how computational methods, augmented by advances in artificial intelligence (AI), have opened new avenues for uncovering the origins and functions of neural timescales, while clarifying their different definitions (Box~1).
We highlight three directions for how computational methods can disentangle distinct types of timescales and synthesize diverse empirical findings into quantitative theories (\fig\ref{fig:intro_schem}):

\begin{itemize}
    \item ``Neural timescales'' are measured from different brain signals, task contexts, and computational methods: how do these measurements relate and differ from one another?
    \item Which cellular and network processes shape measured timescales, and how can mechanistic models of brain dynamics help propose and exclude candidate mechanisms? 
    \item When are ``representational neural timescales'' that mirror environmental timescales necessary (or sufficient) for supporting task-relevant computations, and how to use functional models from machine learning to explore such hypotheses?
\end{itemize}

\begin{tcolorbox}[width=\textwidth,colback={white},title={\textbf{Box 1. Glossary: Different usages of ``timescales''}},colbacktitle={babyblueeyes},coltitle=black, float, floatplacement=!ht]

The definition of timescale varies across studies.
We briefly outline different notions of timescale that are commonly (and sometimes implicitly) used, with more detailed discussions throughout the Review:

\paragraph{Biophysical time constants:} Timescales can correspond to passive membrane or adaptation time constants of single neurons. Similarly, timescales measured from field potential data may relate to time constants of synaptic or ion channel currents (Box~2).

\paragraph{Activity timescales in linear(ized) networks:} Another common notion refers to the decay time constant of population activity. Specifically, the output of a linear time-invariant system (i.e., a linear recurrent neural network) can be decomposed into independent modes of (complex) exponentials with different decay time constants (Box~2). This notion is often leveraged to interpret estimated timescales with the implicit assumption that network dynamics is (locally) linear, since ``time constants'' in nonlinear networks are state-dependent and thus ill-defined as a global, fixed property.

\paragraph{Response timescales:} Aside from the above ``intrinsic'' properties of the neural system under observation, measured timescales can also reflect timescales embedded in the input or be modulated by task and behavioral states. Response timescales could thus correspond to multiple timescales in a streaming sensory stimulus (e.g., a movie), or more generally, to upstream neural activity onto the downstream, recorded region (e.g., V1 to V4). They can also arise from modulations by top-down inputs during a behavioral task. \\

Any quantification of timescales from neural recordings is likely to reflect a mixture of the above. Therefore, a key challenge moving forward for both computational methods and experimental design lies in separating those properties of neural systems from measurements of neural activity, as well as understanding their roles in neural circuit dynamics and computation. 

\end{tcolorbox}

\section{Estimating timescales from neural recordings} 
\label{section:methods}

Neural timescales have been estimated in various experimental paradigms, including different species, tasks, and recording modalities, 
and using different methods (Table~\ref{table:exp_data}). 
Although studies often emphasize convergent results, different paradigms and methods introduce assumptions and biases on estimated timescales that may lead to important conceptual deviations, which
should be incorporated when assessing the generalizability of results.

\begin{figure}[!t]
    \centering
     \includegraphics[trim=0 0 0 0, clip, width = \linewidth]{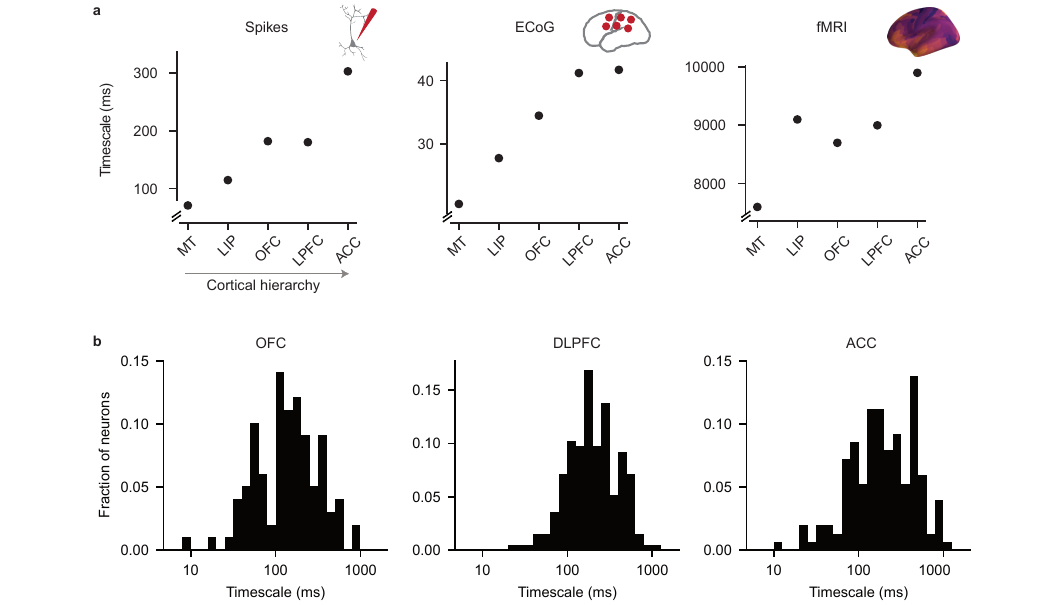}
\caption{\textbf{Diversity of neural timescales across and within brain areas.} 
 \textbf{a.} Average estimated timescales from spontaneous neural activity measured by different recording modalities consistently increase along the cortical hierarchy, but their values differ by orders of magnitude (data from~\cite{murray_hierarchy_2014,gao_neuronal_2020,manea_intrinsic_2022}). \textbf{b.} Neural timescales are also diverse across individual neurons within one brain area (data from~\cite{cavanagh_autocorrelation_2016}). MT: medial-temporal area, LIP: lateral intraparietal area, OFC: orbitofrontal cortex, LPFC: lateral prefrontal cortex, DLPFC: dorsolateral prefrontal cortex, ACC: anterior cingulate cortex. }
    \label{fig:exp_data}
\end{figure}

\subsection{The impact of experimental paradigm on estimated timescales}

\newcommand{\colnhp}[1]{\textcolor{JungleGreen}{#1}}
\newcommand{\colh}[1]{\textcolor{RoyalPurple}{#1}}
\newcommand{\colr}[1]{\textcolor{CarnationPink}{#1}}
\renewcommand{\arraystretch}{1.2} 
\setlength{\extrarowheight}{2pt} 
\begin{table}[!t]
    \fontsize{10}{12}\selectfont 
    \centering
    \renewcommand{\tabularxcolumn}[1]{>{\centering\arraybackslash}m{#1}} 
    \begin{tabularx}{17cm}{|c|>{\centering\arraybackslash}X|>{\centering\arraybackslash}X|>{\centering\arraybackslash}X|>{\centering\arraybackslash}X|}
        \hline
         \diagbox[width=3.4cm, height=1.5cm]{Modality}{Method} & Model-free & Variants of exponential fit & Autoregressive models & Bayesian  methods \\ \hline
        Spiking activity 
        &\colnhp{Mendoza-Halliday et~al.~}\cite{mendoza2014sharp}$^*$, \colnhp{Wang et~al.~}\cite{wang_flexible_2018}$^*$, \colnhp{Meirhaeghe et al.~}\cite{meirhaeghe_precise_2021}$^*$, \colr{Tang et~al.~}\cite{tang2021multiple}$^*$ 
        & \colh{Zeisler et al.~}\cite{zeisler2025consistent}$^{**}$, \colnhp{Murray et al.~}\cite{murray_hierarchy_2014}, \colnhp{Wasmuht et al.~}\cite{wasmuht_intrinsic_2018}, \colnhp{Cavanagh et al.~}\cite{cavanagh_autocorrelation_2016}, \colnhp{Fascianelli et al.~}\cite{fascianelli_autocorrelation_2019}, \colnhp{Cirillo et al.~}\cite{cirillo2018neural}, \colnhp{Bernacchia et al.~}\cite{bernacchia2011reservoir}$^*$, \colnhp{Fontanier et al.~}\cite{fontanier_inhibitory_2022}$^{**}$, \colnhp{Zeisler et al.~}\cite{zeisler2025consistent}$^{**}$, \colr{Siegle et al.~}\cite{siegle_survey_2021}$^{**}$, \colr{Rudelt et al.~}\cite{rudelt_signatures_2023}$^{**}$, \colr{Rudelt et al.~}\cite{rudelt_embedding_2021}$^*$, \colr{Piasini et al.~}
        \cite{piasini2021temporal}$^{**}$, \colr{Zeisler et al.~}\cite{zeisler2025consistent}$^{**}$, 
        \colr{Imani et al.~}\cite{imani_predictive_2023},
        \colr{Shi et al.~}\cite{shi2025brain}
        &
        \colnhp{Spitmaan et al.~}\cite{spitmaan_multiple_2020}$^{**}$, \colnhp{Song et al.~}\cite{song_hierarchical_2023}$^{**}$, \colr{Song et al.~}\cite{song_hierarchical_2023}$^{**}$ 
        & \colnhp{Zeraati et al.~}\cite{zeraati_flexible_2022}, \colnhp{Zeraati et al.~}\cite{zeraati_intrinsic_2022}$^{**}$, \colr{Neophytou et al.~}\cite{neophytou_recurrent_2021}, \colr{Gon{\c{c}}alves et al.~}\cite{gonccalves_training_2020}, \colr{Gao et al.~}\cite{Gao2024automind}, \colr{Vinogradov et al.~}\cite{Vinogradov2024excitability}\\ \hline
        Calcium imaging 
        & \colr{{\c{C}}atal et al.~}\cite{ccatal2024flexibility}$^{**}$ 
        & \colnhp{Runyan et al.~}\cite{runyan_distinct_2017}$^*$,
        \colr{{\c{C}}atal et al.~}\cite{ccatal2024flexibility}$^{**}$,
        \colr{Imani et al.~}\cite{imani_predictive_2023}
        & \colr{Pinto et al.~}\cite{pinto_multiple_2022}$^{**}$ 
        & \colr{Nair et al.~}\cite{nair_line_2023}$^{*}$, \colr{Vinograd et al.~}\cite{Vinograd2024causal}$^{*}$, \colr{Vinogradov et al.~}\cite{Vinogradov2024excitability}\\ \hline
        LFP, ECoG, EEG 
        & \colh{Honey et al.~}\cite{honey_slow_2012}$^*$,
        \colh{{\c{C}}atal et al.~}\cite{ccatal2024flexibility}$^{**}$
        & \colh{Donoghue et al.~}\cite{donoghue_parameterizing_2020}$^{**}$, \colh{Brake et al.~}\cite{brake_aperiodic_2021}, \colh{Chaudhuri et al.~}\cite{chaudhuri_random_2018}, 
        \colh{Gao et al.~}\cite{gao_neuronal_2020}$^{**}$, \colh{{\c{C}}atal et al.~}\cite{ccatal2024flexibility}$^{**}$, \colh{Cusinato et al.~}\cite{cusinato2025sleep}, \colnhp{Gao et al.}\cite{gao_neuronal_2020}, \colnhp{Manea et al.~}\cite{manea_neural_2023}$^{**}$ 
        & 
        & \\ \hline
        fMRI 
        & \colh{Hasson et al.~}\cite{hasson2008hierarchy}$^*$, \colh{Raut et al.~}\cite{raut_hierarchical_2020}, \colh{Fallon et al.~}\cite{fallon_timescales_2020}, \colh{Watanabe et al.~}\cite{watanabe_atypical_2019}, \colh{Shafiei et al.~}\cite{shafiei2020topographic}, \colnhp{Manea et al.~}\cite{manea_intrinsic_2022}, \colr{Sethi et al.~}\cite{sethi2017structural} 
        & \colh{Fallon et al.~}\cite{fallon_timescales_2020} 
        & 
        & \\ \hline
    \end{tabularx}
    \caption{\textbf{Overview of timescale measurements}. Timescales have been measured using various recording modalities, estimation methods, and species (indicated by colors as \colh{human}, \colnhp{non-human primate}, and \colr{rodent}). * indicates studies that measured timescales from task-driven activity, and ** from both spontaneous and task-driven activity. The rest measured timescales from spontaneous activity. Repeated entries indicate multiple methods, modalities, or species.}
    \label{table:exp_data}
\end{table}

Despite consistent hierarchical organization of neural timescales estimated across different modalities (Table~\ref{table:exp_data}), their absolute values differ by orders of magnitude (\fig\ref{fig:exp_data}a).
Spiking activity shows timescales of tens to hundreds of milliseconds~\cite{murray_hierarchy_2014,spitmaan_multiple_2020}, LFP/ECoG estimates are an order of magnitude faster~\cite{manea_neural_2023,gao_neuronal_2020}, while fMRI BOLD signals exhibit much longer timescales on the order of seconds~\cite{raut_hierarchical_2020,fallon_timescales_2020,manea_intrinsic_2022}.
These differences may arise because each modality is dominated by distinct, though related, physiological processes: LFP and ECoG primarily reflect fast transmembrane and synaptic currents, spiking measures neuronal output dynamics, and BOLD captures slower hemodynamic responses coupled to population activity~\cite{logothetis_neurophysiological_2001}, leading to systematically different timescale estimates.
The recording setup can further limit the detection of faster timescales, for instance, due to the delay of the hemodynamic response~\cite{kim1997limitations}, the time constant of calcium release and calcium indicator~\cite{papaioannou2022advantages}, and imaging scan rates.
Thus, when interpreting their potential functional relevance, it is important to distinguish whether estimated timescale differences arise from underlying neurophysiological properties, measurement process constraints, or both.

Neural timescales can also be estimated from different behavioral states and task conditions (Table~\ref{table:exp_data}), resulting in two conceptually different quantities: (i) intrinsic timescales reflecting spontaneous neural activity fluctuations, and (ii) response timescales arising from stimulus- or task-driven activity fluctuations.
While both intrinsic and response timescales increase along the cortical hierarchy~\cite{murray_hierarchy_2014,siegle_survey_2021,rudelt_embedding_2021,piasini2021temporal,manea_intrinsic_2022,gao_neuronal_2020,manea_neural_2023,raut_hierarchical_2020,spitmaan_multiple_2020,song_hierarchical_2023,zeraati2024census,honey_slow_2012,hasson2008hierarchy}, they are not correlated within single neurons~\cite{spitmaan_multiple_2020}.
The reason behind this observation may be two-fold: 
First, single-neuron activity timescales are highly diverse within individual brain areas~\cite{cavanagh_autocorrelation_2016,wasmuht_intrinsic_2018,fascianelli_autocorrelation_2019,cirillo2018neural}, and even areas at the top of the hierarchy have many neurons with fast timescales (\fig\ref{fig:exp_data}b).
Various studies have shown that neurons with fast and slow timescales engage in different task epochs and are relevant to distinct computations~\cite{cavanagh_diversity_2020}. 
Second, neural timescales are state- and input-dependent, and change according to task demands~\cite{gao_neuronal_2020,zeraati_intrinsic_2022,manea_neural_2023,fontanier_inhibitory_2022} and input timescales~\cite{piasini2021temporal}.
Thus, distinct mechanisms may underlie intrinsic and response timescales.
In the following sections, we discuss how generative model-based methods can incorporate modality- and task-specific components when estimating neural timescales and how mechanistic models can disentangle their underlying mechanisms.

~
\subsection{Different methods for estimating timescales}

\begin{figure}[!t]
    \centering
     \includegraphics[trim=0 0 0 0, clip, width = \linewidth]{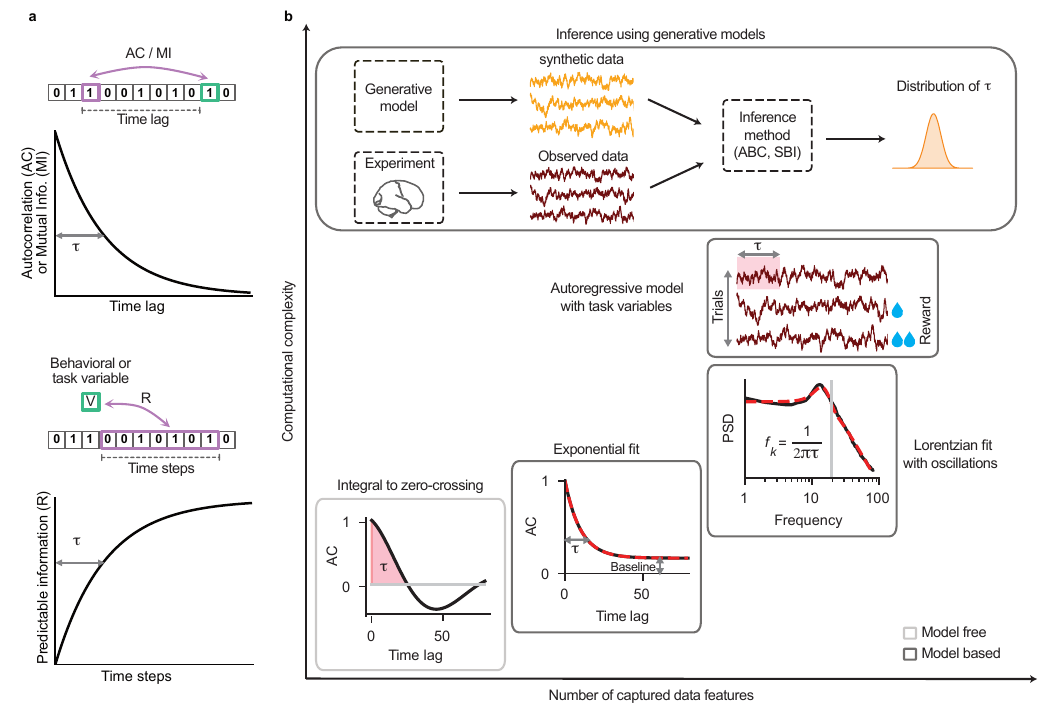}
\caption{\textbf{Different methods for estimating neural timescales.} 
 \textbf{a.} Neural activity fluctuations can be characterized by various summary statistics such as the autocorrelation or delayed mutual information (AC / MI, top), or information content in neural activity about a preceding stimulus (bottom) (adapted from~\cite{runyan_distinct_2017, rudelt_embedding_2021,rudelt_signatures_2023}). \textbf{b.} Methods for estimating timescales ($\tau$) from summary statistics generally vary along two axes: (i) Computational complexity (vertical), (ii) Number of features they capture in data (horizontal): from the approximate decay rate to capturing autocorrelation shape by including a baseline representing slow processes across trials or oscillatory components, to including task-related components and prior knowledge about the data (panels are adapted from~\cite{watanabe_atypical_2019,murray_hierarchy_2014,donoghue_how_2022,spitmaan_multiple_2020,zeraati_flexible_2022}). }
    \label{fig:mod_meth}
\end{figure}

Extracting timescales from neural activity usually requires two steps: (i) computing a summary statistic that quantifies temporal fluctuations, and (ii) estimating timescales from the summary statistic (\fig\ref{fig:mod_meth}). 
For the summary statistic, most methods use some variant of time-lagged metrics, such as autocorrelation (AC, or its frequency domain equivalent, power spectral density, PSD) or delayed mutual information (MI, \fig\ref{fig:mod_meth}a, top).
For example, intrinsic timescales can be characterized by the decay rate of the spontaneous activity autocorrelation computed from individual experimental trials~\cite{murray_hierarchy_2014,cavanagh_autocorrelation_2016,wasmuht_intrinsic_2018,fascianelli_autocorrelation_2019,bernacchia2011reservoir,fontanier_inhibitory_2022,siegle_survey_2021,rudelt_signatures_2023,zeraati_intrinsic_2022,zeraati_flexible_2022,gao_neuronal_2020,imani_predictive_2023,piasini2021temporal,donoghue_parameterizing_2020,chaudhuri_random_2018,ccatal2024flexibility,manea_intrinsic_2022,manea_neural_2023,fallon_timescales_2020,raut_hierarchical_2020,watanabe_atypical_2019}, whereas response timescales are quantified by the autocorrelation of trial-averaged activity post-stimulus~\cite{piasini2021temporal,siegle_survey_2021}.
Response timescales can also be estimated with decoding methods that assess how long the encoding of a given stimulus or task event persists over time~\cite{runyan_distinct_2017,rudelt_embedding_2021,rudelt_signatures_2023,piasini2021temporal}.
Here, the summary statistic is often constructed using metrics such as decoding accuracy or predictability as a function of time (\fig\ref{fig:mod_meth}a, bottom)~\cite{runyan_distinct_2017,rudelt_embedding_2021,rudelt_signatures_2023}.

Then, using the computed summary statistic (e.g., AC or MI) as input, methods for estimating timescales rely on distinct theoretical assumptions and incorporate different features of neural dynamics.
Here, we catalog existing methods along two axes: (i) computational complexity (often as a result of using more detailed models) and (ii) the number of features they capture in data (\fig\ref{fig:mod_meth}b).
While we mainly focus here on quantifying timescales from the autocorrelation function, similar analyses can be applied to information-theoretic summary statistics.

The first group of methods does not assume any strict mathematical form for the autocorrelation shape. These methods characterize the decay rate as, for example, the time lag where autocorrelation drops to $1/e$~\cite{beiran_contrasting_2019} or 0.5~\cite{raut_hierarchical_2020}), or the integral of positive values~\cite{watanabe_atypical_2019,manea_intrinsic_2022}. We refer to these methods as ``model-free''.
Computing such measures is straightforward, estimated directly from the data.
However, such model-free methods do not consider the full shape of data autocorrelation. As a result, in the presence of complex temporal features in neural activity beyond a single exponential decay,, such as oscillatory modes, different methods cancan potentially give different results on the same data (e.g.,~\cite{beiran_contrasting_2019,Muscinelli2019-ct}).

The second group of methods assumes a mathematical function for the autocorrelation to capture the relevant data features (i.e., model-based), and estimate timescales from the parameters of the fitted function.
The simplest assumption is that the autocorrelation of neural activity follows an exponential decay function, $\textrm{AC}(t) = e^{-t/\tau}$, where decay time constant $\tau$ reflects the timescale of underlying dynamics (which is equivalent to a Lorentzian function in the frequency domain, i.e., $\textrm{PSD}(f) \sim (f^2 + f_k^2) ^{-1}$, $\tau = (2\pi f_k)^{-1}$~\cite{gao_neuronal_2020}).
The exponential decay can be further augmented with additional features to better capture the complexity of temporal dynamics, such as modeling multiple timescales by a linear mixture of exponential functions~\cite{chaudhuri_random_2018,rudelt_signatures_2023,shi2025brain} or very slow dynamics by adding an offset~\cite{murray_hierarchy_2014,cavanagh_autocorrelation_2016,spitzner_mr_2021}.

Neural activity can additionally contain oscillatory dynamics that appear as periodic fluctuations in the autocorrelation or peaks in the PSD, which can impact the estimated timescale.
To overcome this problem, spectral parameterization methods~\cite{donoghue_parameterizing_2020,gao_neuronal_2020, brake_aperiodic_2021} model oscillatory peaks as a mixture of Gaussians, which are removed before fitting an extended form of the Lorentzian function.

Finally, a more accurate estimate of timescales can be obtained using a generative process that captures the statistics of neural time series~\cite{zeraati_flexible_2022,spitmaan_multiple_2020,neophytou_recurrent_2021,strey_estimation_2019}.
Such methods are particularly powerful since they can explicitly incorporate modality-specific properties and task variables.
For instance, including task variables in an autoregressive generative process has allowed for disentangling intrinsic and response timescales~\cite{spitmaan_multiple_2020,pinto_multiple_2022,song_hierarchical_2023}.
Moreover, multiple timescales can be explicitly modeled in the generative process, for instance, as a mixture of autoregressive processes with distinct timescales~\cite{zeraati_flexible_2022,zeraati_intrinsic_2022,spitmaan_multiple_2020}. 
Generative models can also capture experimental limitations, such as short trial durations that affect the shape of data autocorrelation and can lead to systematic underestimation of timescales by model-free and model-based methods discussed earlier~\cite{zeraati_flexible_2022}.

The parameters of the generative process can be directly fitted to neural time series~\cite{spitmaan_multiple_2020} (e.g., via maximum likelihood estimation) or can be used to instantiate a generative model in Bayesian inference methods~\cite{zeraati_flexible_2022,strey_estimation_2019, neophytou_recurrent_2021,gonccalves_training_2020, deistler2025simulation, Boelts2025reloaded, Gao2024automind, linderman2017bayesian}.
Bayesian methods aim to uncover the entire set of parameters compatible with the data (i.e., the posterior distribution) rather than obtaining the single best-fitting value of timescales (or other parameters), allowing them to better deal with the stochasticity of neural activity.
For simple models, the posterior distribution can be computed analytically~\cite{strey_estimation_2019}, but for more complex models, simulation-based methods such as approximate Bayesian computation (e.g., abcTau~\cite{zeraati_flexible_2022}) and simulation-based inference (SBI~\cite{gonccalves_training_2020,deistler2025simulation,Boelts2025reloaded, Cranmer2020sbi,Gao2024automind}) can be applied to obtain approximate posterior distributions.
In such methods, synthetic time series are generated from an explicit generative model mimicking similar statistics as neural recordings. 
Then, by matching the summary statistics of synthetic and experimental data, we can estimate timescales using the parameters of the matched generative model.
The generative model and summary statistics can be chosen flexibly to explicitly capture different data features (e.g., spiking activity as a point process versus LFP as a continuous signal~\cite{kass_analysis_2014,zhou_establishing_2015,safavi_univariate_2021}, or particular biophysical properties capturing different signal modalities).
Therefore, while these methods often carry additional computational costs from performing simulations and posterior inference, they can incorporate important mechanistic assumptions---both based on neurobiology and biophysics of the signal-generating process---when estimating timescales.

\section{Mechanistic models of timescales in the brain} 
\label{section:mechanistic}
\begin{figure}[!t]
    \centering
     \includegraphics[trim=0 0 0 0, clip, width = \linewidth]{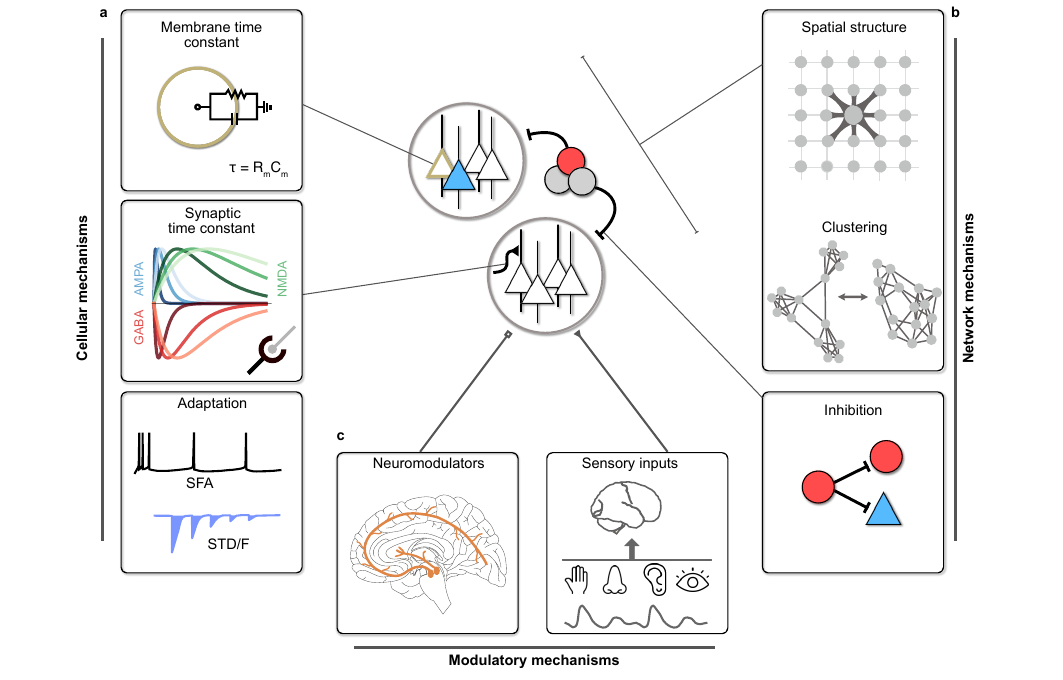}
\caption{\textbf{Mechanisms underlying generation and modulation of neural timescales.} 
  Computational models show that neural timescales are shaped by cellular biophysical properties (a, e.g., membrane, synaptic, and adaptation time constants) and network interactions (b, e.g., spatially structured or clustered connectivity, inhibition). Additionally, timescales can be modulated by neuromodulatory signals and sensory inputs from the environment (c). SFA: spike-frequency adaptation; STD/F: short-term synaptic depression/facilitation.}
    \label{fig:mechanisms}
\end{figure}

Once measured, mechanistic models constrained by empirical timescales can illustrate how diverse timescales emerge within and across brain areas. Such models have been used to account for (i) the hierarchical organization of timescales across areas \cite{chaudhuri_large-scale_2015}, (ii) their diversity within a single area \cite{chaudhuri_diversity_2014,stern_reservoir_2021,zeraati_intrinsic_2022}, and (iii) their flexibility under different task demands \cite{zeraati_intrinsic_2022,wang_flexible_2018,beiran_parametric_2021,trepka_training_2024}. 
Proposed mechanisms span cellular and synaptic processes and network interactions, with bottom-up and top-down inputs, and neuromodulation further shaping timescales in stimulus- and task-dependent manners (\fig\ref{fig:mechanisms}).

\subsection{Cellular and synaptic mechanisms}
\label{sec:model_cellular}
In computational models, neural timescales are often set by biophysical time constants that capture heterogeneous ionic and synaptic properties ~\cite{gjorgjieva_computational_2016,duarte_synaptic_2017}. 
Most dynamical system models assign each neuron a membrane time constant (typically 5–50 ms~\cite{azouz_cellular_1999}; Box~2), while slower timescales can be implemented by additional mechanisms such as slow synaptic kinetics~\cite{harish_asynchronous_2015,beiran_contrasting_2019,clark2024theory}, short-term synaptic plasticity~\cite{mongillo2008synaptic,Rolls2016adaptation}, or adaptation~\cite{brette_adex_2005,carter_cannabinoid-mediated_2007,pozzorini_whitening_2013,beiran_contrasting_2019}. 
Spike-frequency adaptation, in particular, spans a wide spectrum of fast and slow components~\cite{la2006multiple}, potentially following a power-law distribution~\cite{pozzorini_whitening_2013}. Such diverse adaptation timescales have been linked to efficient coding~\cite{pozzorini_whitening_2013} and improved processing of temporally dispersed inputs~\cite{salaj_spike_2021}.

On the other hand, it is difficult to disentangle the exact contribution of these biophysical time constants to the autocorrelation timescales of measured neural activity---henceforth referred to as \emph{activity timescale} (Box~2).
For instance, although activity timescales generally increase with longer synaptic time constants~\cite{beiran_contrasting_2019}, their dependence on the adaptation time constant may be weaker~\cite{beiran_contrasting_2019} and potentially more complex~\cite{Muscinelli2019-ct} in a network.
These observations support the notion that cellular biophysics affect neural activity differently when modulated by a network than in isolation~\cite{fontanier_inhibitory_2022}.
Nevertheless, even in models where network connectivity alone can explain heterogeneous activity timescales, introducing additional biophysical details (e.g., H-current) can provide better quantitative fits to data~\cite{borst2025differential}.

\subsection{Network and connectivity mechanisms}
\label{sec:model_network}

Neurons are embedded in rich networks~\cite{bullmore_complex_2009,buxhoeveden_minicolumn_2002} that shape the correlation structure of their activity \cite{smith_spatial_2013,safavi_nonmonotonic_2018,rosenbaum_spatial_2017,shi_cortical_2022}.
Therefore, a mechanistic understanding of activity timescales requires the context of interactions within a network in conjunction with cellular biophysics.
Most models rely on recurrent (excitatory or inhibitory) connectivity to explain how diverse neural timescales emerge, but as we discuss later, feedforward networks with rich temporal inputs can also give rise to diverse and hierarchical timescales~\cite{wiskott2002slow}.

Recurrent connectivity (within or across brain areas) can create positive or negative feedback that controls activity timescales. 
This can be done, for example, through creating attractor dynamics that give rise to persistent activity (e.g., line attractors)~\cite{seung_how_1996, Goldman_2003_line, wang2024bifurcation} or shifting the network dynamical state toward a critical transition~\cite{kadmon_transition_2015,wilting_operating_2018,chaudhuri_random_2018,zeraati_intrinsic_2022,zeraati2024topology, sooter2024cortex, meisel2017decline,yellin2023modulation,wardak2022extended}.
By analyzing linear recurrent networks, we can define and compute the activity timescales of neurons analytically based on the eigenvalues of the connectivity matrix (Box~2).
However, in nonlinear networks, such timescales are not uniquely defined and can only be specified for a given (linearized) dynamical state, such as steady-state dynamics around fixed points~\cite{wang2024bifurcation,maheswaranathan2019reverse,singh2023emergent,nair_line_2023,Vinograd2024causal}.

\begin{tcolorbox}[width=\textwidth,colback={white},title={\textbf{Box 2. Activity timescale vs. biophysical time constants}},colbacktitle={babyblueeyes},coltitle=black]
\label{box:timeconstant}
``Neural timescales'' in the literature, for the most part, refer to the estimated autocorrelation decay timescale of measured neural activity, i.e., by describing the autocorrelation function as $\textrm{ACF}(t)=e^{\frac{-t}{\tau_\textrm{fit}}}$. Methods discussed in Section \ref{section:methods} generally aim to find the value of $\tau_\textrm{fit}$ (or multiple $\tau_\textrm{fit} $s) that best fit the data.\\

In contrast, cellular (e.g., Hodgkin-Huxley) and network (e.g., RNN) models in Sections \ref{section:mechanistic} and \ref{section:functional} typically include \textit{time constant} parameters that describe how the impact of input into a neuron decays over time, e.g., $\tau_\textrm{param}\frac{dV}{dt} = -(V(t)-V_\textrm{rest}) + I(t)$, where $I(t)$ includes time-varying bottom-up (stimulus-driven), top-down, and recurrent inputs.\\

It is important to note that the former, estimated $\tau_\textrm{fit}$ does not necessarily correspond to the latter, underlying model parameter $\tau_\textrm{param}$, and in general, does not. Section \ref{section:mechanistic} outlines multiple mechanisms, including those unrelated to time constants, such as network connectivity, that contribute to shaping $\tau_\textrm{fit}$ we observe in data.\\

As an illustrative example, take the activity timescale of a linear recurrent network as
\[
\frac{dx(t)}{dt} = Ax(t)
\]
where \(x(t)\) is the state vector, e.g., neural activity at time \(t\), and \(A\) is a square connectivity matrix.
To acquire the timescales of the system dynamics, we solve for the eigenvalues of the matrix \(A\), \(\lambda_i\), through the characteristic equation, \(A\textbf{u}_i = \lambda_i \textbf{u}_i\). \\

In general, eigenvalues are complex and have the form \(\lambda_i = \textrm{Re}(\lambda_i) + j\textrm{Im}(\lambda_i)\). For a stable decaying mode, the real part of the associated eigenvalue must be negative, and its decay timescale is defined as
\begin{equation*}
\tau_i = -\frac{1}{\textrm{Re}(\lambda_i)}.
\end{equation*}
The dominant activity timescale of the system (i.e., \(\tau_{\textrm{fit}}\)) is then typically determined by the mode whose eigenvalue has the largest real part. Furthermore, for a linear network composed of units with fixed membrane time constant \(\tau_\textrm{param}\), the system dynamics follow
\begin{equation*}
\tau_{param}\frac{dx(t)}{dt} = Ax(t).
\end{equation*}
Assuming \(\tau_\textrm{param}\) to be the same for all neurons, the activity timescale is scaled accordingly,
\begin{equation*}
\tau_i = -\frac{\tau_\textrm{param}}{\textrm{Re}(\lambda_i)}.
\end{equation*}
This approach has been applied to connectivity matrices incorporating further biological constraints, e.g., Dale's law~\cite{rajan_eigenvalue_2006}. For nonlinear networks, it is necessary to first linearize the system around a specific point before conducting the above analysis (see, e.g., \cite{singh2023emergent}).
\end{tcolorbox}

The strength and structure of recurrent excitation influence the amount of positive feedback in a network and, consequently, the autocorrelation timescale of neural activity. 
Timescales generally increase with the strength of excitation~\cite{ostojic_two_2014,chaudhuri_large-scale_2015,van_meegen_microscopic_2021,hart_recurrent_2020,demirtas_hierarchical_2019,kadmon_transition_2015,wilting_operating_2018,chaudhuri_large-scale_2015,chaudhuri_random_2018,zeraati_intrinsic_2022}, and heterogeneity in connection strengths can generate distinct timescales across units \cite{chaudhuri_diversity_2014,stern_reservoir_2021,chen2024searching,wardak2022extended,bernacchia2011reservoir}.
Even with fixed connection strengths, the connectivity structure (i.e., topology) strongly influences temporal dynamics. 
Clustered networks can produce metastable states whose noise-driven transitions generate slow timescales \cite{litwin-kumar_slow_2012,Aljadeff2015-hi,huang_once_2017}, a phenomenon also seen in homeostatically regulated spiking networks \cite{cramer_autocorrelations_2022}. 
More generally, spatially structured connectivity introduces multiple dominant timescales, yielding autocorrelation shapes that deviate from simple exponential decay \cite{shi_spatial_2022,zeraati_intrinsic_2022}. 
In such networks, interactions at larger spatial scales give rise to slower timescales, consistent with theories and observations on the spatiotemporal organization of oscillatory brain activity \cite{buzsaki_neuronal_2004}.

Negative feedback through recurrent inhibition also contributes to shaping neural timescales. 
Strong inhibitory-to-inhibitory connections can generate the slow timescales necessary for working memory in recurrent network models \cite{kim_strong_2021}. 
In detailed biophysical models of excitatory and inhibitory neurons, inhibition regulates the stability and transitions of metastable states, producing slow timescales \cite{fontanier_inhibitory_2022}. 
In these models, the full autocorrelation profile of frontal activity is captured by a combination of cellular mechanisms (e.g., after-hyperpolarization potassium currents and GABA-B conductances) and network interactions. 
Furthermore, recurrent inhibition, together with membrane and synaptic time constants, can explain the distinct temporal dynamics observed in spiking activity and LFP signals \cite{safavi_uncovering_2023}.

\subsection{Input-driven and modulatory mechanisms for dynamically shaping timescales}

The mechanisms discussed above explain how specific ranges of timescales arise from ``static" properties of single neurons and networks. 
However, neural timescales are dynamic and can adapt to task demands~\cite{gao_neuronal_2020,zeraati_flexible_2022,fontanier_inhibitory_2022,wang_flexible_2018,manea_neural_2023,trepka_training_2024,piasini2021temporal,beiran_parametric_2021}.
Here, we discuss how external inputs (bottom-up and top-down) and neuromodulators alter neural timescales.

External inputs can modulate both intrinsic and response timescales.
Incoming input can alter biophysical properties, such as bringing individual neurons to a high-conductance state (lower resistance), which reduces the membrane time constant~\cite{pare_impact_1998} to facilitate the distinction of distant synaptic inputs~\cite{destexhe_high_2003}.
The spatiotemporal statistics of the input can shape response timescales, as different stimulus features vary on distinct timescales (e.g., transient motion versus stable object identity). 
Feedforward models constructed based on slow feature analysis predict that fast input dynamics are represented at lower levels of the visual hierarchy, while slower features emerge at higher levels \cite{wiskott2002slow,berkes2005slow,franzius2007slowness,franzius2011invariant,halvagal2023combination, matteucci2024unsupervised}, yielding progressively more stable representations. 
These predictions have been validated experimentally in the rodent visual system \cite{piasini2021temporal}. 
Similar principles apply in recurrent networks, where populations with fast (slow) intrinsic timescales preferentially respond to fast (slow) input components \cite{stern_reservoir_2021}. 
Furthermore, stimulus spatial scale modulates population timescales by shaping feedforward and recurrent interactions \cite{litwin-kumar_spatial_2012}. 
Thus, depending on the spatiotemporal profile of the input, different timescales may be encoded and reflected in neural activity.

Inputs can also modulate activity timescales indirectly by shifting the dynamical state of nonlinear networks.
In nonlinear networks, timescales are only defined at a specific state.
Different input levels thus place the network in different states, which can alter the ``local'' timescales~\cite{shi_spatial_2022,wang2024bifurcation}.
This mechanism could explain the modulation of neural timescales during selective-attention~\cite{zeraati_intrinsic_2022}, aggression~\cite{nair_line_2023}, and timing tasks~\cite{wang_flexible_2018,beiran_parametric_2021}.
Moreover, it explains how the organization of brain-wide response timescales during tasks deviate from the expected hierarchy of intrinsic timescales~\cite{wang2024bifurcation}.
An example is the sharp transition in response timescales from the middle temporal (MT) area to the medial superior temporal (MST) during the delay period of a working memory task, despite their closeness in the visual hierarchy~\cite{mendoza2014sharp}.
Large-scale cortical models suggest that this transition arises from nonlinear interactions between different brain areas, where distinct intrinsic and delay-activity response timescales are shaped by dynamics in different fixed points~\cite{wang2024bifurcation}.

Neuromodulatory projections from diffuse subcortical nuclei can also alter timescales as a result of their impact on cellular and network properties~\cite{marder_neuromodulation_2012, shine_computational_2021}.
Acetylcholine can change synaptic time constants and lead to persistent firing (e.g., during pain processing~\cite{zhang_metabotropic_2010}).
An increase in acetylcholine, such as during selective attention, can also strengthen the thalamocortical synaptic efficacy by affecting nicotinic receptors, while weakening lateral cortical interactions by affecting muscarinic receptors~\cite{deco_cholinergic_2011}.
Similarly, the different contribution of glutamatergic receptors in feedforward and recurrent pathways affects the time course of neural population responses~\cite{self_different_2012}. 
Thus, neuromodulators can indirectly adapt neural timescales to task demands through selective gain modulation.

Altogether, a combination of cellular, network, input, and neuromodulatory mechanisms can shape the diversity and flexibility of neural timescales.
We argue that constraining mechanistic computational models with realistic timescales can provide new insights into circuit mechanisms underlying neural dynamics.
While such models are typically challenging to constrain, recent works have leveraged model-based estimation methods (as discussed above) to discover data-consistent mechanistic models~\cite{gonccalves_training_2020, gao_gbi_2023, Gao2024automind, Vinogradov2024excitability, Doorn2024-oz}.
Based on the available experimental measurements, one can constrain models with a single estimated timescale (e.g., average timescale of a brain area in \fig\ref{fig:exp_data}a), or with a distribution of timescales (e.g., diverse timescales within an area \fig\ref{fig:exp_data}b).
As we discuss next, assessing for realistic timescales (e.g., similar to realistic firing rates) allows us to test the plausibility of different models aimed to explain computation and function.

\section{Timescale as signatures of dynamics supporting network computation}
\label{section:functional} 

Finally, returning to function: Recent studies have shown that single-unit and population timescales correlate with task performance and behavioral timescales~\cite{cavanagh_diversity_2020,soltani_timescales_2021,wolff_intrinsic_2022}. 
However, establishing causal contributions of specific timescales will require precise control over network dynamics and computation, which remains challenging in experiments but is an area where computational models can offer essential complementary insights.

\begin{figure}[!t]
    \centering
     \includegraphics[trim=0 0 0 0, clip, width = \linewidth]{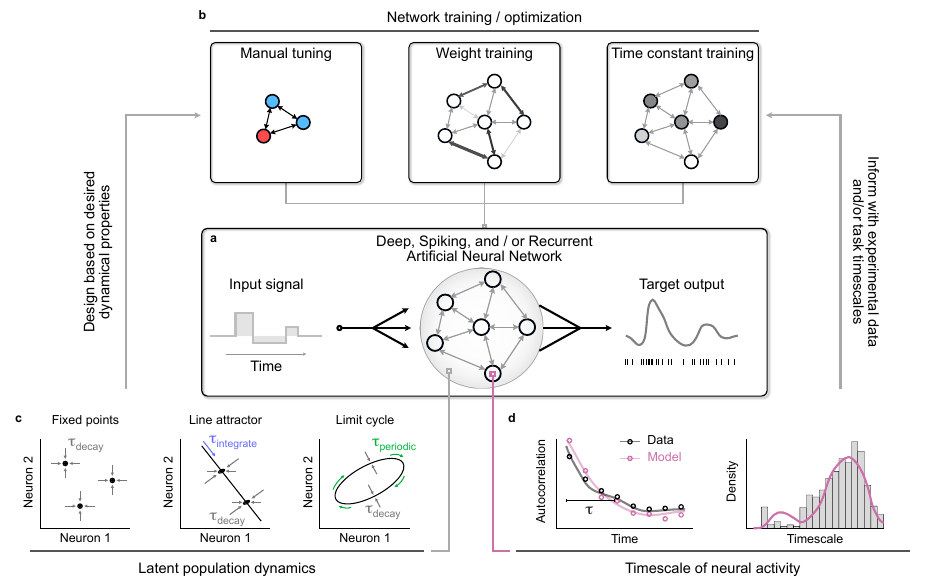}
\caption{\textbf{Task-performing models reveal functional relevance of timescales.} 
\textbf{a.} Network models are set up to achieve input-output transformations similar to tasks performed by animals and humans.
\textbf{b.} Model parameters, such as network weights and neuronal time constants, can be manually tuned when feasible or computationally optimized via gradient descent.
\textbf{c.} A certain geometry of population dynamics is desirable for network computation, such as fixed points, line attractors, and limit cycles.
\textbf{d.} Activity timescales and fitted time constants in network models match experimental data.
}
    \label{fig:functional}
\end{figure}

Task-performing artificial neural networks (ANN)---including spiking and rate-based recurrent neural networks (SNNs and RNNs), and deep neural networks---offer viable testbeds.
These models are designed to mimic the input-output behavior of animals performing a task~\cite{Richards2019-yh,Yang2020-ih} (\fig\ref{fig:functional}a).
To achieve this, network parameters are either ``hand-crafted'' when possible (\fig\ref{fig:functional}b, left), or ``trained'' using an optimization algorithm (\fig\ref{fig:functional}b, middle).
Networks can then be analyzed and perturbed to generate hypotheses about computational mechanisms, particularly with respect to the geometry (Fig.~\ref{fig:functional}c)~\cite{Sussillo2014-ra} and timescales of latent dynamics (\fig\ref{fig:functional}d).

Here, we discuss two categories of investigations: First, we review works that study timescales as an emergent characteristic of task-performing networks, including classical models with ``hand-tuned'' dynamical properties and task-optimized networks. Second, we discuss recent studies that directly optimize time constants and timescale-related parameters to enhance computational capacity (\fig\ref{fig:functional}b, right) in biologically plausible SNNs and RNNs, and in deep learning models.

\subsection{Emergent timescales in task-performing networks}

Computational neuroscience has a long-standing interest in the repertoire of network dynamics---and by extension, their timescales---to elucidate how neural circuits perform computations with rich temporal dependencies.
In classical hand-tuned models, fixed points, line attractors, and limit cycles (\fig\ref{fig:functional}c) are dynamical motifs that can be leveraged for memory~\cite{hopfield_hopfield_1982, lim_balanced_2013}, integration and decision-making~\cite{seung_stability_2000, xjwang_slow_2002, wong_recurrent_2006}, and timing~\cite{buonomano_timing_2003}, respectively (~\cite{khona_attractor_2022}).
Beyond their geometry, the timescale of latent dynamical systems near these low-dimensional manifolds are also key properties of interest. 
Crucially, those timescales govern how quickly and stably networks can integrate new input while remembering or forgetting old ones, filter out noisy fluctuations to sense slower signals, and multiplex inputs of different frequencies.
Consistent with hand-tuned models~\cite{vogels_neural_2005}, recent task- and activity-optimized models also exhibit fixed points corresponding to persistent inputs or memories~\cite{mante_context-dependent_2013}, line attractors for tracking slow environmental or behavioral variables~\cite{nair_line_2023,Vinograd2024causal, Liu2024mating, Mountoufaris2024neuropeptide}, and (quasi-)limit cycles traversed at different speeds corresponding to the timing of inputs or outputs~\cite{wang_flexible_2018, beiran_parametric_2021}.

Not only is the distribution of timescales important for implementing computations, they are useful for identifying dynamical structures in the first place. For instance, an asymmetry in the decay time constants of a (locally) linear system (Box~2) is indicative of one slow direction and many fast directions---a line attractor~\cite{Maheswaranathan2019universality, nair_line_2023}. Identification of such structures in data-constrained models further enables causal manipulation of the populations involved, closing the loop between theory and experiment in dissecting circuit mechanisms underlying complex behaviors, such as the role of subcortical structures in encoding affect~\cite{Vinograd2024causal, Liu2024mating, Mountoufaris2024neuropeptide}. 

Related research on reservoir computing~\cite{maass_liquidstate_2004, jaeger_harnessing_2004} similarly considered network timescales from the perspective of latent dynamical systems, as they require long-range but non-divergent dynamics at the edge-of-chaos~\cite{bertschinger_realtime_2004, boedecker_chaos_2012}.
Since reservoir networks typically had trained input and output weights but fixed recurrent weights, much attention was dedicated to investigating how they can be initialized to optimize computational capacity~\cite{ozturk2007ESN, Steiner2023clusterinit}, and in particular, how input-driven chaos and a repertoire of timescales (or temporal filters) can be leveraged for time-dependent computations~\cite{bertschinger_realtime_2004}.

Alternatively, one can directly compare single-neuron activity timescales with neurons in task-performing models (\fig\ref{fig:functional}d) to probe network computations, such as temporal scaling~\cite{wang_flexible_2018} and working memory~\cite{kim_strong_2021, khajehabdollahi_emergent_2024}. 
This approach can be further extended to study more complex networks, such as those with more biologically-realistic constraints like Dale's law~\cite{song_training_2016,kim_strong_2021} and dendritic compartmentalization~\cite{brenner_tractable_2022}, hypothesized computational mechanisms such as oscillatory coding~\cite{Liebe2025phase, pals_phaselock_2023}, as well as better analytical tractability (e.g., low-rank RNNs~\cite{Mastrogiuseppe2018-zy}), improved expressiveness, and trainability~\cite{jordan_gated_2021, krishnamurthy_theory_2022, park_persistent_2023}. Targeting the same sought-after dynamical repertoires, ``hybrid'' approaches can leverage different weight initialization and regularization strategies to nudge networks toward line attractors or limit cycles after training~\cite{schmidt_nonlinear_2019, park_persistent_2023}.

Altogether, these studies provide insights into dynamical computations in specific tasks and network architectures. Since the geometry of neural representation is intimately linked to its dynamics (and their timescales), and the latter is similarly critical for computation~\cite{Ostrow2023-ex}, both classical and recent works lend support to the importance of diverse timescales for different computations. In other words, the timescales of dynamics are signatures of the phase portrait geometry, and constraining one to satisfy certain requirements often constrains the other as well~\cite{krishnamurthy_theory_2022}. Nevertheless, it may be possible for networks to acquire similar geometries but vastly different timescales. Therefore, we suggest explicit comparisons of \textit{in silico} and \textit{in vivo} timescales in task-performing networks as a straightforward metric to shed light on the dynamics involved. It should be noted that similar to nonlinear biophysical networks (Section~\ref{sec:model_network}) analyses of timescales in task-performing models also rest on the assumption of linearized systems. While such analyses provide insights into ``local'' dynamics and computations, individual timescales do not inform us of global network properties, which may require sampling the distribution of timescales across multiple states and task contexts.

\subsection{Optimization of timescale-related parameters}\label{sec:optimize_timescales}

Another powerful way to study the role of timescales on computation \textit{in silico} is to directly optimize related model parameters, such as synaptic or adaptation time constants, or phenomenological variables like temporal receptive field size. 

In standard RNNs, all neurons have a single fixed time constant, which determines how quickly the effect of an input decays. In reality, neurons in a network have a range of membrane time constants, due to varying properties such as membrane capacitance (Section \ref{sec:model_cellular}). As such, single-neuron time constants can be individually optimized as parameters~\cite{tallec_chrono_2018,perez-nieves_neural_2021, fang2021incorporating, Lappalainen2024connectome,stefanidi_pretraining_2023,khajehabdollahi_emergent_2024}, or even parameterized to be input-dependent (e.g., long short-term memory units, LSTMs). Thus, by comparing to networks not endowed with heterogeneous time constants, such studies provide evidence for the potential necessity of diverse timescales in computations:

Several recent studies have shown that task-trained recurrent and feedforward spiking networks can benefit from optimization of heterogeneous single-neuron time constants~\cite{kim_strong_2021,perez-nieves_neural_2021, salaj_spike_2021,fang2021incorporating, Moro2024hierarchy}. Strikingly, trained networks exhibit timescale distributions similar to those found in the real brain ~\cite{perez-nieves_neural_2021, kim_strong_2021}. Notably, time constants do not necessarily need to match task timescales, but both longer and more diverse time constants were found to be beneficial~\cite{salaj_spike_2021}. Similarly, RNNs with trainable time constants demonstrate improved task performance and memory capacity while the learned time constants recover timescales in the data-generating process~\cite{Quax2020-az, heinrich_adaptive_2018}. With more biologically-realistic models, such as connectome-constrained networks, training both the connection weights and single-neuron time constants resulted in emergent stimulus tuning similar to that of \emph{in vivo} recorded neurons~\cite{Lappalainen2024connectome}. 

Remarkably, multi-layer feed-forward DNNs and RNNs trained to mimic the input-output relationship of a single morphologically accurate pyramidal neuron acquire complex and spatially varying temporal filters~\cite{Beniaguev2021-ck}, emphasizing the computational capacity provided by multiple timescales even at the scale of single neurons~\cite{spieler2024the}.
On the other hand, network-mediated timescales may be a more robust mechanism for performing long memory tasks rather than trainable time constants~\cite{khajehabdollahi_emergent_2024}, further highlighting the need to disambiguate the computational roles of cellular and network-mediated timescales.

These functional benefits may extend to deep learning models. Instead of trainable or fixed time constants, they can be further made to be input-dependent via gating parameters (similar to gated RNNs and LSTMs), which improves modeling of sequences with multiple timescales~\cite{heinrich_learning_2020}. Furthermore, initializing gating parameters to reflect the distribution of timescales in the data improves performance on tasks with long sequences~\cite{tallec_chrono_2018}. Leveraging this observation, endowing LSTM units with a distribution of effective timescales constructed to match the power-law temporal structure of natural language improves network performance in language modeling, especially long-range dependencies~\cite{mahto2021multitimescale}. In parallel, these ``chrono initialized" LSTMs improved decoding of brain response along the cortical hierarchy in humans~\cite{Jain2020-sq}, potentially unifying mechanisms in artificial and biological neural networks. 

Finally, recent research in deep learning has shown a convergence of principles based on thoughtful inductive biases and initialization, as well as self-supervised pretraining, for learning tasks with long temporal dependencies.
Although RNN-based architectures can model infinitely long dependencies in theory, simpler implementations based on finite-length temporal convolutions appeared to be more effective at acquiring long timescale memories~\cite{bai2018empirical}, though both architectures learn effective temporal context windows that reflect natural structures in the data (e.g., speech~\cite{Keshishian2021-rj}).

More recent architectures based on deep linear state space models (SSMs, e.g., LMU~\cite{Voelker2019LMU}, S4~\cite{Gu2021-qo}, and others~\cite{Gu2021-LSS, Gu2023-ma, Orvieto2023resurrecting, Hasani2022liquid}) enjoy the benefits of both recurrence and convolutions, outperforming Transformer-based architectures on long timescale tasks, which are hindered by quadratic complexity in time.
Remarkably, the success of linear SSM-based models was in part driven by replacing the standard dense recurrent connections with complex diagonal matrices, which directly parameterize the eigenvalues of a linear system---i.e., the system timescales (Box~2).
Later extensions capitalize on these insights to derive better parameterizations and initializations with simpler implementations ~\cite{li2023sgconv, Orvieto2023resurrecting}.
On the other hand, both Transformer-based and SSM-based architectures can be drastically improved through self-supervised pretraining on the input sequences before downstream supervised learning on the same long-range tasks~\cite{tay2021lra, amos2024never}.
These findings suggest that pretraining or directly equipping models with inductive biases appropriate for task-relevant timescales can be promising, but the exact mechanisms of these improvements---and to what extent they are related to intrinsic timescales in biological networks---remain unclear.

In summary, there is growing interest in convergent approaches to produce task-relevant emergent dynamics and timescales in network models, as well as explicit optimization of time constant parameters. Leveraging these approaches, findings from biological and deep learning models suggest that both the magnitude and diversity of timescales may be important. Thus, these works represent promising collaborative approaches in neuroscience and AI aimed towards understanding and improving time-dependent computations in biological and artificial systems.

\section{Discussion} 

Computational approaches provide a complementary framework for translating empirical observations of diverse neural timescales into quantitative, testable hypotheses. While different estimation methods can disentangle various components of neural activity, mechanistic models and task-performing artificial neural networks can uncover circuit mechanisms underlying brain dynamics and computation. Thus, combining these approaches could provide a principled understanding of how the brain integrates information across multiple timescales to support perception, learning, and action in complex, dynamically changing environments.

Much of systems and cognitive neuroscience has focused on a narrow temporal window defined by common trial-based task structures and recording techniques. 
Expanding this window to include fast synaptic processes, slow physiological modulations, and open-ended naturalistic behavior will be essential for understanding how timescales interact across orders of magnitude.
Achieving this goal will require new experimental paradigms, long-term recordings, and modeling approaches explicitly designed to investigate neural dynamics across multiple orders of magnitude in time.

Finally, advances in machine learning and AI offer new opportunities to link neural timescales with principles of dynamic computation. Viewing network timescales as signatures of the underlying dynamical landscape suggests that shaping timescale repertoires may be an effective way to optimize both biological and artificial networks for computation, learning, and stability. In this context, constraining artificial neural networks to match empirically observed activity timescales provides a promising route toward models that not only perform tasks but also exhibit realistic neural dynamics. Conversely, modern AI systems, capable of operating over long and naturalistic timescales, enable the study of temporal computation beyond standard laboratory paradigms. Together, these developments position neural timescales as a unifying axis for integrating data, theory, and models, and for advancing our understanding of how the brain supports adaptive behavior in a dynamic world.

\subsection*{Acknowledgments}
This work was supported by the German Research Foundation (DFG) through Germany’s Excellence Strategy (EXC-Number 2064/1, PN 390727645) (RZ, JHM, RG) and SFB1233 (PN 276693517) (JHM),   the Sofja Kovalevskaja Award from the Alexander von Humboldt Foundation endowed by the Federal Ministry of Education and Research (RZ, AL), the Max Planck Society (RZ), the European Union (ERC, ``DeepCoMechTome'', ref. 101089288) (JHM), the European Union’s Horizon 2020 research and innovation program under the Marie Skłodowska-Curie grant agreement No.~101030918 (AutoMIND) (RG) and an add-on fellowship from the Joachim Herz Foundation (RZ).

We thank Ana Manea and Jan Zimmermann for providing the timescales of fMRI data (\fig\ref{fig:exp_data}a), Mattia Chini, Bradley Voytek, and the speakers and participants of the Cosyne 2022 workshop on \href{https://www.rdgao.com/blog/2022/03/28/}{``Mechanisms, functions, and methods for diversity of neuronal and network timescales"} for discussions.

We acknowledge the drawings and icons provided by \href{https://scidraw.io/}{SciDraw} (the brain in \fig\ref{fig:intro_schem}, \ref{fig:mod_meth}, and \ref{fig:mechanisms}b, sensory adapted from \href{https://doi.org/10.5281/zenodo.3925981}{https://doi.org/10.5281/zenodo.3925981}, the brain in \fig\ref{fig:mechanisms}b, neuromodulators adapted from \href{https://doi.org/10.5281/zenodo.4312476}{https://doi.org/10.5281/zenodo.4312476}, the pyramidal neuron in \fig\ref{fig:intro_schem} and \ref{fig:exp_data} adapted from \href{https://doi.org/10.5281/zenodo.3925905}{https://doi.org/10.5281/zenodo.3925905}, and the water drop in \fig\ref{fig:mod_meth} adapted from \href{doi.org/10.5281/zenodo.3925935}{doi.org/10.5281/zenodo.3925935}), 
 and Pixel Perfect from \href{https://www.flaticon.com/}{Flaticon} (\href{https://www.flaticon.com/free-icon/open-hands_3179232?term=hand&page=1&position=22&origin=style&related_id=3179232}{hand}, \href{https://www.flaticon.com/free-icon/nose_4012060?term=nose&page=1&position=4&origin=style&related_id=4012060}{nose}, \href{https://www.flaticon.com/free-icon/ear_4011593?term=ear&page=1&position=5&origin=style&related_id=4011593}{ear}, and \href{https://www.flaticon.com/free-icon/sight_4011607?term=sight&page=1&position=2&origin=style&related_id=4011607}{eye} icons in \fig\ref{fig:mechanisms}).

\subsection*{Author contributions}
RZ and RG conceptualized the overall structure of the review, wrote the first draft, and prepared the figures. 
All authors discussed, revised, and wrote the final version of the paper.

\subsection*{Competing interests}
The authors declare no competing interests.

\nolinenumbers
\bibliographystyle{naturemag} 
\bibliography{refs.bib} 
\end{document}